\documentclass[aps,prx,twocolumn,longbibliography,superscriptaddress,floatfix,nofootinbib]{revtex4-2}
\usepackage{epsfig,amsmath,amssymb,color,comment,physics}
\usepackage[makeroom]{cancel}
\usepackage[caption=false]{subfig}
\usepackage{mathrsfs}
\usepackage[countmax]{subfloat}
\usepackage[normalem]{ulem}
\usepackage[english]{babel}
\setcitestyle{super}
\usepackage{dsfont}
\usepackage{float}
\usepackage[bookmarks=true,colorlinks,linkcolor=black,urlcolor=NavyBlue,citecolor=RoyalBlue]{hyperref}
\usepackage{mathtools}
\usepackage{siunitx}
\usepackage{upgreek}
\usepackage{comment}
\usepackage{xcolor}

\hypersetup{
    unicode=false,     
    pdftoolbar=false,  
    pdfmenubar=true,   
    pdffitwindow=false, 
    pdfstartview={FitH},
    pdftitle={},    
    pdfauthor={Authors},     
    pdfsubject={},   
    pdfcreator={},   
    pdfproducer={}, 
    pdfkeywords={quantum many-body scars} {superconducting processor} {quantum state tomography}, 
    pdfnewwindow=true,
    colorlinks=true,
    linkcolor=black,
    citecolor=blue, 
    filecolor=magenta,
    urlcolor=blue
}

\newcommand{\beginsupplement}{%
    \setcounter{table}{0}
    \renewcommand{\thetable}{S\arabic{table}}%
    \setcounter{figure}{0}
    \renewcommand{\thefigure}{S\arabic{figure}}%
    \setcounter{equation}{0}
    \renewcommand{\theequation}{S\arabic{equation}}%
    \setcounter{section}{0}
    \renewcommand{\thesection}{S\arabic{section}}%
   }

\newcommand{\sign}{\text{sgn}}

\newcommand{\Er}{E_{\textrm{r}}}

\newcommand{\micron}{\upmu\textrm{m}}

\begin{document}

\title{Anyonization of bosons}

\author{ Sudipta  Dhar } \thanks{These authors contributed equally to this work.}
\affiliation{Institut f{\"u}r Experimentalphysik und Zentrum f{\"u}r Quantenphysik, Universit{\"a}t Innsbruck, Technikerstra{\ss}e 25, Innsbruck, 6020, Austria} 

\author{ Botao Wang } \thanks{These authors contributed equally to this work.}
\affiliation{Center for Nonlinear Phenomena and Complex Systems,
Université Libre de Bruxelles, CP 231, Campus Plaine, B-1050 Brussels, Belgium}

\author{ Milena  Horvath} \thanks{These authors contributed equally to this work.}
\affiliation{Institut f{\"u}r Experimentalphysik und Zentrum f{\"u}r Quantenphysik, Universit{\"a}t Innsbruck, Technikerstra{\ss}e 25, Innsbruck, 6020, Austria} 

\author{ Amit Vashisht }
\affiliation{Center for Nonlinear Phenomena and Complex Systems,
Université Libre de Bruxelles, CP 231, Campus Plaine, B-1050 Brussels, Belgium}

\author{ Yi Zeng }
\affiliation{Institut f{\"u}r Experimentalphysik und Zentrum f{\"u}r Quantenphysik, Universit{\"a}t Innsbruck, Technikerstra{\ss}e 25, Innsbruck, 6020, Austria}

\author{ Mikhail B. Zvonarev }
\affiliation{Université Paris-Saclay, CNRS, LPTMS, 91405 Orsay, France}

\author{Nathan Goldman}
\affiliation{Center for Nonlinear Phenomena and Complex Systems,
Université Libre de Bruxelles, CP 231, Campus Plaine, B-1050 Brussels, Belgium}
\affiliation{Laboratoire Kastler Brossel, Collège de France, CNRS, ENS-Université PSL,
Sorbonne Université, 11 Place Marcelin Berthelot, 75005 Paris, France}

\author{ Yanliang  Guo }\email{yanliang.guo@uibk.ac.at}
\affiliation{Institut f{\"u}r Experimentalphysik und Zentrum f{\"u}r Quantenphysik, Universit{\"a}t Innsbruck, Technikerstra{\ss}e 25, Innsbruck, 6020, Austria}

\author{ Manuele  Landini}\email{manuele.landini@uibk.ac.at}
\affiliation{Institut f{\"u}r Experimentalphysik und Zentrum f{\"u}r Quantenphysik, Universit{\"a}t Innsbruck, Technikerstra{\ss}e 25, Innsbruck, 6020, Austria}

\author{ Hanns-Christoph  N{\"a}gerl}\email{christoph.naegerl@uibk.ac.at}
\affiliation{Institut f{\"u}r Experimentalphysik und Zentrum f{\"u}r Quantenphysik, Universit{\"a}t Innsbruck, Technikerstra{\ss}e 25, Innsbruck, 6020, Austria}

\date{\today}

\begin{abstract}
Anyons~\cite{Leinaas1977,Wilczek1982} are low-dimensional quasiparticles that obey fractional statistics, hence interpolating between bosons and fermions. In two dimensions, they exist as elementary excitations of fractional quantum Hall states~\cite{Tsui1982,Laughlin1983,Wilczek1984,Halperin1984,STERN2008} and they are believed to enable topological quantum computing~\cite{Kitaev2003,RevModPhys.80.1083}. One-dimensional (1D) anyons have been theoretically proposed, but their experimental realization has proven to be difficult. Here, we observe anyonic correlations, which emerge through the phenomenon of spin-charge separation~\cite{ogata1990,Vijayan2020,Hulet2022}, in a 1D strongly-interacting quantum gas. The required spin degree of freedom is provided by a mobile impurity, whose effective anyonic correlations are associated with an experimentally tunable statistical angle. These anyonic correlations are measured by monitoring the impurity momentum distribution, whose asymmetric feature demonstrates the transmutation of bosons via anyons to fermions~\cite{Calabrese2007,Patu2007,Santachiara2007,Axel2015,hao1}. Going beyond equilibrium conditions, we study the dynamical properties of the anyonic correlations via dynamical fermionization of the anyons~\cite{delcampo}. Our work opens up the door to the exploration of non-equilibrium anyonic phenomena in a highly controllable setting~\cite{patu1,wright,Wang2014,Piroli2017,Hao3,delcampo,Liu2018}.
\end{abstract}
\maketitle

Quantum theory tells us that particles can be categorized into two distinct groups based on the phase $\theta$ that the quantum wavefunction accumulates when two particles are exchanged~\cite{Dirac1930-DIRTPO}. 
This phase is crucial to the collective behavior of ensembles of identical particles: bosonic particles, with $\theta\!=\!0$, may pile up and condense into the same state, whereas fermions, with $\theta\!=\!\pi$, follow Pauli's exclusion principle and avoid each other. This has drastic consequences, e.g., forming the basis for the table of elements and assuring stability of neutron stars in the case of fermions, and giving rise to spectacular phenomena such as superfluidity, superconductivity, and laser emission for bosons. But in dimensions lower than three, more exotic possibilities exist. In the seminal works by Leinaas and Myrheim~\cite{Leinaas1977}, and Wilczek~\cite{Wilczek1982}, it was realized that a new type of particle, called anyon, with arbitrary values of $\theta$ is possible. Anyons behave neither as bosons nor as fermions. They obey fractional quantum statistics~\cite{Haldane1991} and are expected to show an intermediate correlation behavior, interpolating between bosons and fermions.

Two-dimensional anyons are found to exist as quasiparticles in topological states of matter, such as fractional quantum Hall states in solid-state systems~\cite{Bartolomei2020,Nakamura2020,Feldman2021}, and they can be engineered in superconducting quantum processors~\cite{2021Satzinger,2023Google,2024Xu}, Rydberg atom arrays~\cite{2021Semeghini}, and trapped-ion processors~\cite{2024Iqbal}. Triggered by Haldane's fractional exclusion statistics~\cite{Haldane1991}, anyons in 1D have attracted a lot of theoretical attention. A wealth of phenomena has been proposed, such as statistically induced phase transitions and fractional Mott insulators~\cite{Keilmann2011}, anomalously bound pairs~\cite{Greschner2018}, the accumulation of Friedel oscillations with increasing $\theta$~\cite{Eckardt}, and dynamical fermionization and bosonization of anyons~\cite{delcampo,wright}. Anyonic models in 1D have been studied both in the continuum~\cite{kundu1999,Bonkhoff2021} and on discrete lattices~\cite{Keilmann2011,Santos}. As a hallmark for the presence of anyonic correlations, an asymmetric momentum distribution~\cite{Axel2015,hao1,hao2} is expected. The theoretical underpinnings of 1D anyons have long intrigued the scientific community, yet their experimental realization and the observation of their dynamical behavior have remained elusive. Recently, using a Floquet drive, 1D anyons have been realized in a two-atom lattice setting~\cite{Greiner_anyon}

\begin{figure*}
\centering
\renewcommand{\figurename}{Fig}
\includegraphics[width=\linewidth]{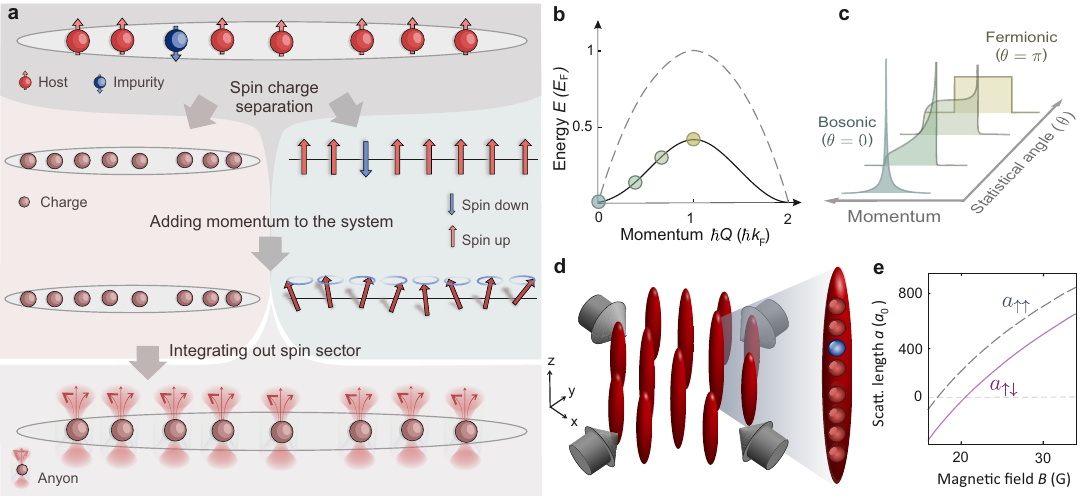}
\caption{{\bf Experimental realization of 1D anyons.} {\bf{a}}, Illustration of the emergence of anyons from spin-charge separation. For strong interactions in 1D, the wavefunction factorizes into a charge part and a spin part. In the finite momentum ground state of the system, all the momentum is carried by the spin sector in the form of spin waves. Integrating out the spin degrees of freedom realizes an effective system of 1D hardcore anyons in the charge sector. The statistical phase $\theta$ of these emerging anyons is given by the momentum of the spin waves, see Methods.
{\bf{b}}, Edge of the excitation spectrum of a 1D Bose gas for charge excitation (dashed line) and spin excitation (solid line)~\cite{meinert2017bloch}. 
{\bf{c}}, Expected momentum distributions of anyons~\cite{Gamayun2020} for different values of the statistical angle $\theta$ as set by the momentum $\hbar Q$ and indicated in {\bf{b}}.
{\bf{d}}, Experimental realization of an ensemble of 1D Bose gases in tubes formed by two retro-reflected laser beams. Each tube contains on average one impurity particle (blue sphere). It can be made to strongly interact with the strongly-correlated TG host gas (red spheres). {\bf{e}}, Host-host (dashed curve) and host-impurity (solid curve) scattering lengths $a_{\uparrow \uparrow}$ and $a_{\uparrow \downarrow}$ as a function of the magnetic field $B$.}
\label{fig1}
\end{figure*}

Here, we present a cold-atom realization of a many-body system with anyonic correlations in 1D. We use a degenerate gas of strongly interacting Cs atoms to simulate 1D hardcore anyons with an arbitrary statistical phase $\theta$. Our system consists of a single spin impurity embedded in and strongly interacting with a Tonks-Girardeau (TG) host gas. The impurity serves a dual purpose in our study: it enables the generation of anyonic correlations in the system and acts as a probe to observe these correlations. 

For strong impurity-host interaction, spin-charge separation occurs in our system~\cite{ogata1990,giamarchi2003quantum}, see Fig.~\ref{fig1}a, with the many-body wavefunction of $N$ particles factorizing into a spatial $\varphi(x_1, x_2, .., x_N )$ and a spin part $\chi(\sigma_1, \sigma_2, .., \sigma_N )$, where $\sigma_i =\uparrow, \downarrow$ is the spin of the $i$-th particle.
Anyonic correlations on $\varphi$ arise from engineering a spin wavefunction with fractional exchange symmetry when we restrict to cyclic permutations. For this, the spin wavefunction is prepared in eigenstates of the cyclic spin permutation operator $\hat{C}$, i.e., spin waves $|\theta\rangle$ with eigenvalue $e^{-i\theta}$, see Methods. We experimentally prepare the spin wave by adiabatically accelerating the impurity along the low-energy edge of the excitation spectrum to momentum $\hbar Q$, see Fig.~\ref{fig1}b. In each particular state, the momentum of the spin wave fixes the effective phase shift resulting from an exchange of the impurity with one of the particles in the host gas. For $Q\!=\!0$, the exchange results in no phase shift, akin to bosonic statistics, while for $Q\!=\!k_\text{F}$, the resulting phase shift is $\pi$ as expected from fermionic statistics~\cite{SM}. For intermediate momenta, we expect that anyonic statistics is realized. Here, $k_\text{F}\!=\!\rho\pi$ denotes the Fermi momentum of the TG gas, with $\rho$ the 1D density.

A particular observable that is sensitive to the anyonic correlations and to the statistical phase $\theta$ is the momentum distribution of the impurity. Specifically, the one-body correlators of the impurity and for a hardcore anyon system are equal~\cite{PhysRevA.109.012209,Pu2017},
\begin{equation}
    (\langle\varphi|\otimes\langle\theta|)\hat{b}_{ \downarrow}^\dagger(x)\hat{b}_{\downarrow}(y)(|\theta\rangle\otimes|\varphi\rangle)=\frac{1}{N}\langle\varphi|\hat{a}^\dagger(x) \hat{a}(y)|\varphi\rangle,
    \label{momentumdistribution}
\end{equation} 
where $\hat{b}^\dagger_{\downarrow}$ ($\hat{b}_{\downarrow}$) is the creation (annihilation) field operator of the impurity and $\hat{a}^\dagger$ ($\hat{a}$) is the anyon creation (annihilation) field operator, with $(\hat{a}^\dagger)^2\!=\!0$ defining the hardcore condition. Equation (\ref{momentumdistribution}) gives us direct access to the anyonic momentum distribution. Figure~\ref{fig1}c illustrates the expected anyonic momentum distribution. As $\theta$ is varied, the evolution from a bosonic via a skewed to a fermionic distribution can clearly be seen.

\begin{figure*}
    \renewcommand{\figurename}{Fig.}
    \includegraphics[width=\linewidth]{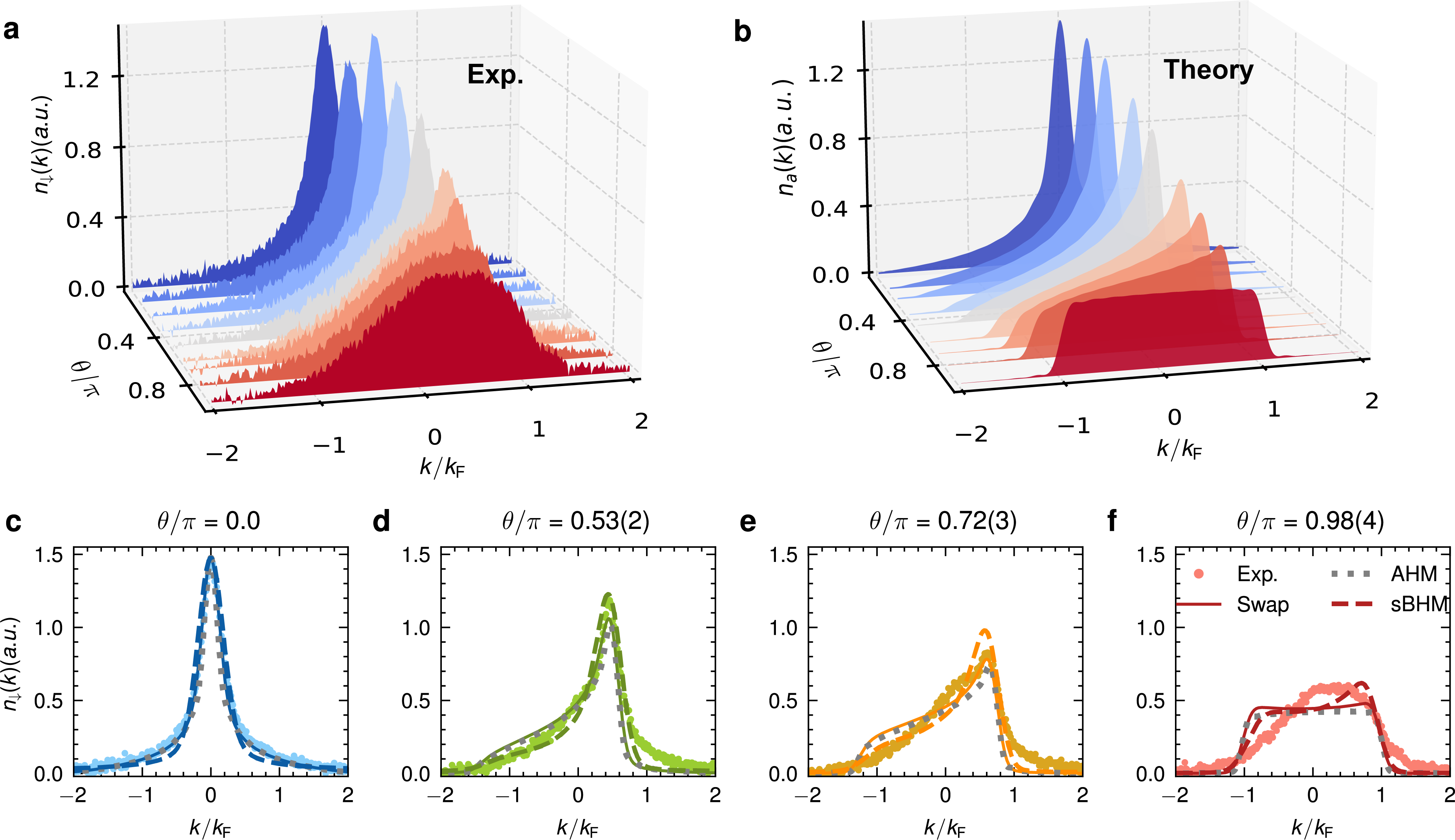}
    \caption{{\bf{Momentum distribution of anyonized bosons.}} {\bf{a}}, Evolution of the measured $n_\downarrow(k)$ for variable exchange phases  $\theta$, determined by the injected momentum $\hbar Q$, as indicated. Each distribution is the average of $7$ experimental realizations. {\bf{b}}, Numerical results of the anyonic momentum distribution $n_a(k)$ using the anyon-Hubbard model. 
    {\bf{c}}-{\bf{f}}, Example distributions for $\theta/\pi$ equal to 0, 0.53(2), 0.72(3), 0.98(4), respectively. The error bars are smaller than the symbol sizes. The data is compared to numerical results of the ground states of the anyon-Hubbard model (dotted lines), the swap model (solid lines) and the time evolution governed by the spinful Bose-Hubbard model (dashed lines).}
    \label{fig2}
\end{figure*}

In the experiment, we prepare an array of about 6000 vertically oriented 1D Bose gases with a weighted average of 37(2) atoms by loading a weakly interacting 3D Bose-Einstein condensate of Cs atoms~\cite{weber2003bose} into a 2D optical lattice as illustrated in Fig.~\ref{fig1}d. Initially, all the atoms are in the hyperfine state $|F,m_{F}\rangle\!=\!|3,3\rangle$, which we denote by $|\!\uparrow \rangle$. A magnetic force levitates the atoms against gravity. We then tune the 1D interaction strength $g_\uparrow \propto a_{\uparrow \uparrow}$ by means of a Feshbach resonance for the scattering length $a_{\uparrow \uparrow}$ (see Fig.~\ref{fig1}e) to bring the 1D Bose gases into the TG regime~\cite{kinoshita2004,haller2009}, setting the Lieb-Liniger parameter to $\gamma_{ \uparrow \uparrow}\!\approx\!14$, see Methods. A short radio-frequency pulse generates spin impurities in $|3,2\rangle\!\equiv\!|\!\downarrow \rangle$ out of the host gas. On average, we create one impurity per 1D Bose gas, with the number set by the power and duration of the pulse. We handshake from pure magnetic levitation to a combination of magnetic and optical levitation to allow for a comparatively small force of $F_\downarrow\!\approx\!mg/18$ on the impurities, while the host gas remains fully levitated. Here, $m$ is the mass of Cs atoms and $g$ is the gravitational acceleration. A small force is needed to ensure that the impurity adiabatically follows the lower edge of the excitation spectrum. During the evolution, the impurity experiences a host-impurity interaction strength of $\gamma_{\uparrow \downarrow}\!\approx\!9$, as set by $a_{\uparrow \downarrow}$, see Fig.~\ref{fig1}e and the Methods. Applying $F_\downarrow$ for a variable evolution time $\tau$ places the system at momentum $\hbar Q\!=\!F_\downarrow \tau$. Varying $\tau$ from zero to $3.5$ ms sets the phase $\theta\!=\!Q/\rho\!=\!\pi Q/k_{\text{F}}$ to values between zero and $\pi$. We finally measure the $|\!\downarrow \rangle$-momentum distribution $n_\downarrow(k)$ by switching $\gamma_{\uparrow \downarrow}$ to zero and imaging the $|\!\downarrow\rangle$ atoms after Stern-Gerlach separation and free time-of-flight (TOF) expansion. The results are presented in Fig.~\ref{fig2}a. For $\theta\!=\!0$, the impurity exhibits a momentum distribution $n_\downarrow(k)$ that is symmetric and sharply peaked at momentum $\hbar k\!=\!0$. As $\theta$ is increased towards $\pi$, the distribution skews and the peak gradually disappears as the distribution broadens and flattens. At $\theta\!=\!\pi$ the distribution is flat-top, nearly filling the entire Brillouin zone from $-k_{\text{F}}$ to $k_{\text{F}}$. In essence, the distribution evolves from a bosonic to a fermionic distribution with significant skewness in between.

\begin{figure}[t!]
\renewcommand{\figurename}{Fig.}
	\includegraphics[width=\linewidth]{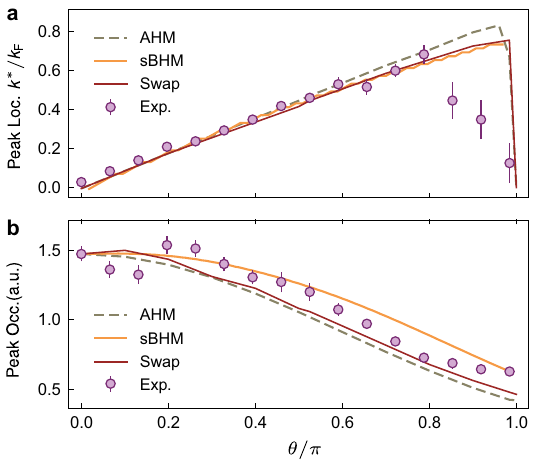}
    \caption{\textbf{Characterization of the anyonic momentum distribution.} {\bf{a}}, measured peak momentum $k^*$ and {\bf{b}}, peak occupation of $n_\downarrow(k)$ as a function of $\theta$. The experimental data (dots) is compared to the results of the simulations based on the various models as indicated. The error bars reflect the standard error.  
}
\label{Fig3}
\end{figure}

\begin{figure*}[t!]
    \renewcommand{\figurename}{Fig.}
    \includegraphics[width=\linewidth]{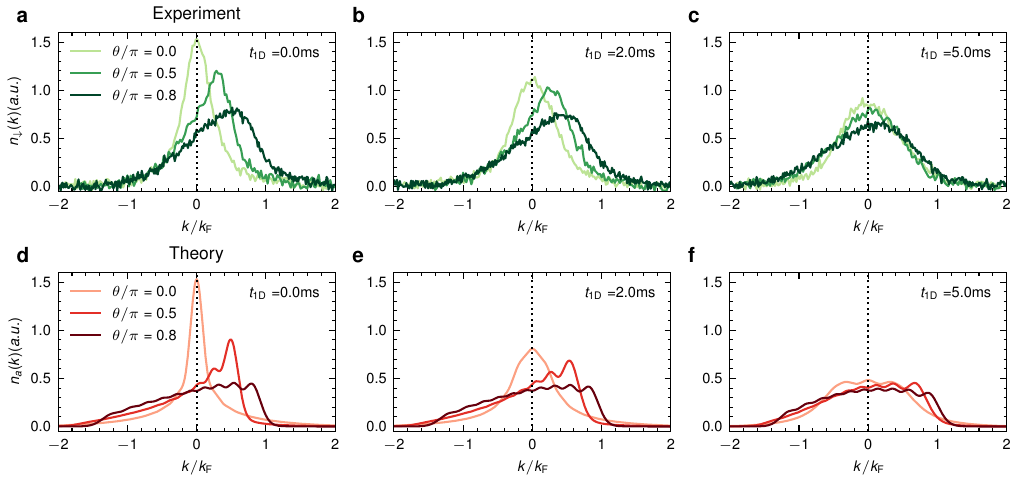}
    \caption{{\bf{Dynamical fermionization of hardcore anyons.}} {\bf{a}}-{\bf{c}}, Evolution of the observed $n_\downarrow(k)$ after quenching the confinement to a flat-bottom trap and allowing 1D expansion for $t_{\rm 1D}=$ 0, 2, 5 ms, for three different $\theta$ as indicated. Each distribution is the average of $10$ experimental realizations. {\bf{d}}-{\bf{f}}, Theoretical prediction of hard-core anyons ($N\!=\!10$) in continuum during 1D free expansion~\cite{delcampo}. 
}
    \label{fig4}
\end{figure*}

The anyonic nature of such skewness behavior is confirmed by performing a quantitative analysis using several theoretical models. Our system is naturally described by a spinful Lieb-Liniger gas, for which an exact Bethe-ansatz solution is available~\cite{McGuire1965,2012Guan} in the limit of a fermionized host gas and which allows an exact anyonic mapping of the impurity momentum distribution in the thermodynamic and hardcore limit~\cite{Gamayun2020,PhysRevA.109.012209}. To properly capture finite-size effects and directly compare the theoretical prediction to the data, we make use of lattice models that we expect to reliably describe the system in the low lattice-filling regime, see Methods. We first turn to the anyon-Hubbard model (AHM)
\begin{equation}
    \hat{H}_{\text{AHM}}=-J\sum_{\ell} \hat{a}^\dagger_\ell \hat{a}_{\ell+1}+h.c.+\frac{U}{2}\sum_{\ell}\hat{n}_\ell(\hat{n}_\ell-1)
    \label{AHM}
\end{equation}
in the hardcore limit with the on-site interaction $U\!\rightarrow\!\infty$. Here, $J$ is the tunneling amplitude between nearest-neighboring sites, $\hat{a}_\ell$ is the anyonic annihilation operator at site $\ell$, and $\hat{n}_\ell=\hat{a}^\dagger_\ell \hat{a}_\ell$ is the particle number operator. The prediction for the momentum distribution as calculated by a matrix product-state algorithm~\cite{2022_itensor,2022_itensor03} is shown in Fig.~\ref{fig2}b. The transition from a peaked bosonic distribution via a skewed to a box-like fermionic distribution can clearly be seen. A direct comparison with our data for selected values of $\theta$ is presented in Fig.~\ref{fig2}(c-f), and we find reasonably good agreement. The second model we employ is the spinful Bose-Hubbard model (sBHM), aimed at describing the dynamics of a spinful Bose system when a force is applied. Additionally, we introduce a novel model, termed the swap model~\cite{SM,Swap}, featuring swap interactions, whose ground state takes the form of the spin wave that we are targeting. The predictions from these models for a single-tube realization are included in Fig.~\ref{fig2}(c-f). Given our finite momentum resolution that results from the inhomogeneous tube distribution (see Methods) of about $0.4 \hbar k_{\text{F}}$, these predictions agree well with our experimental data. \\

The anyonic correlations are reflected in the asymmetric momentum distribution, exhibiting a shift of the peak of the momentum distribution and a variation of the peak value. In Fig.~\ref{Fig3}a we compare the observed peak position $k^*$ and the peak value $n_\downarrow(k\!=\!k^{*})$ with the calculated behavior of an anyonic system as $\theta$ is varied. For small $\theta$, the peak momentum is proportional to $\theta$. The slope of the linear dependence is expected to be proportional to the density $\rho$ of the gas~\cite{Axel2015}. For large $\theta$ close to $\pi$, $k^*$ sharply decreases, since $n_\downarrow(k)$ starts to transmute into a fermionic distribution. Simultaneously, the peak occupation $n_\downarrow(k=k^{*})$ decreases as $\theta$ increases, as shown in Fig.~\ref{Fig3}b. These observations agree well with the results of the numerical calculations. \\

We next turn to the dynamical properties of our anyonized system. Specifically, we perform a rapidity measurement~\cite{Rigol2005,Minguzzi2005}, as has been used recently to study dynamical fermionization in TG gases~\cite{Wilson2020,Li2023Rapidity}. The rapidities are the integrals of motion in a 1D system, and they are expected to follow a fermionic distribution in the hard-core limit also for anyons~\cite{delcampo}. As before, we prepare the system at momentum $\hbar Q$. We then set the force to $F_{\downarrow}\!=\!0$ and the atoms are allowed to expand in an approximately flat potential in 1D by partially compensating the longitudinal harmonic confinement by means of a horizontally propagating blue-detuned anti-trapping laser beam~\cite{Guo_MBDL}. We then measure $n_\downarrow(k)$ as before, but this time after a variable expansion time $t_\text{1D}$ in the 1D tubes. This procedure maps the rapidities onto momentum when $t_\text{1D}$ is chosen to be long enough, typically $5$ ms for our parameters, set by the longitudinal trap frequency and the average particle number (see Methods). The experimental results are shown in Fig.~\ref{fig4}(a-c). In the case without 1D expansion ($t_\text{1D}\!=\!0$), as shown in Fig.~\ref{fig4}a, the initial anyonic momentum distributions $n_\downarrow(k)$ for various values of $\theta$ differ greatly and exhibit the skewness behavior as discussed above. However, as $t_\text{1D}$ is increased, the distributions converge to a similar asymptotic form. At the $t_\text{1D}\!=\!2$ ms (Fig.~\ref{fig4}b) the distributions still differ, but at $t_\text{1D}\!=\!5$ ms (Fig.~\ref{fig4}c) they have become equal. In particular, they have lost their skewness. This behavior is qualitatively captured by our numerical modeling, as shown in Fig.~\ref{fig4}(d-e). For this, starting with anyonic wavefunctions, we simulate the quench dynamics for $N\!=\!10$ anyons after suddenly releasing the harmonic trapping potential by solving the time-dependent Schr\"odinger equation (see Methods). The evolution from greatly differing distributions to nearly symmetric and identical distributions can clearly be seen. Note that the expected shape of the distribution in the long-time limit is set by the harmonic trapping potential~\cite{delcampo,Pu2021,PatuDF2023}. Only for box-shaped trapping one expects a box-shaped distribution. Future experiments with custom-shaped potentials will be able to probe this relationship.

In summary, we have realized a many-body system of 1D anyonized bosons with arbitrary statistical angle from a strongly interacting spinful bosonic system. Our approach relies on the intrinsic fractionalization of spin and charge degrees of freedom in 1D systems in the presence of strong interactions. The observed asymmetric momentum distributions, a hallmark of anyonic correlations, are in good agreement with theoretical predictions. Our findings demonstrate the ability to transmute between bosonic and fermionic behaviors by continuously varying the statistical parameter, thus creating a flexible system that allows for the exploration of anyonic behavior in a controlled, low-dimensional environment. Moreover, the observed phenomenon of dynamical fermionization following a trap quench highlights the complex non-equilibrium dynamics that these systems can exhibit, providing insights into the interplay between quantum statistics and dynamical properties of 1D anyons.

A promising direction for future research will be the realization of tunable interactions between anyons~\cite{Santos2016,Batchelor2006}. This will open up possibilities for the study of exotic quantum phases~\cite{Lange2017,Greschner2017} and phase transitions predicted for 1D anyonic systems~\cite{Keilmann2011,bonkhoff2024}. Our way of realizing density-dependent statistical angles provides a new opportunity to study intriguing dynamical phenomena due to the presence of a statistical interface~\cite{2022Lau}. Generalizing our work beyond 1D to study topologically non-trivial states of matter is also an interesting avenue. Furthermore, our method of measuring non-local string-type correlators via impurities could be used to probe topological order in generic many-body systems~\cite{Grusdt2016,Jain2014,Rougerie2016,Lewenstein2020}.  \\   

\noindent{\bf Acknowledgments}\\
We thank Philip Zechmann and Michael Knap for discussions and providing us at the early stage of this work  with the results of simulations to understand the role of a finite force. We thank Xi-Wen Guan for insightful discussion about fractional statistics of an impurity immersed in spin polarized Fermi gases. \textbf{Funding:} The Innsbruck team acknowledges funding by a Wittgenstein prize grant under Austrian Science Fund (FWF) project number Z336-N36, by the European Research Council (ERC) under project number 789017, by an FFG infrastructure grant with project number FO999896041, and by the FWF's COE 1 and quantA. MH thanks the doctoral school ALM for hospitality, with funding from the FWF under the project number W1259-N27. Work in Brussels is supported by the ERC (LATIS project), the EOS project CHEQS, the FRS-FNRS Belgium and the Fondation ULB. Computational resources have been provided by the Consortium des Équipements de Calcul Intensif (CÉCI), funded by the Fonds de la Recherche Scientifique de Belgique (F.R.S.-FNRS) under Grant No. 2.5020.11 and by the Walloon Region. \textbf{Author Contributions:} The work was conceived by H.C.N., S.D., M.L., M.B.Z. Experiments were prepared and performed by S.D., M.H. Data were analyzed by S.D., B.W., Y.G. Numerical simulations were performed by B.W. ,A.V. Theoretical models were developed by B.W., A.V., N.G., M.B.Z., S.D. and M.L. The manuscript was drafted mainly by H.C.N., M.L., S.D., B.W., Y.G. All authors contributed to the discussion and finalization of the manuscript.
{\bf Data Availability:} The data shown in the main text is available via Zenodo~\cite{Zenodo}.

\bibliography{ref,mybib}

\begin{thebibliography}{82}
\bibitem [{Zen()}]{Zenodo}%
  \BibitemOpen
  \bibinfo {note} {{Data set is available from {Z}enodo at doi: 10.5281/zenodo.14567996}}\BibitemShut {NoStop}%
\end{thebibliography}

\section{Methods}\label{methods}
\subsection{\bf Experiment}\label{expmethods}
The experiment starts with an interaction-tunable 3D Bose-Einstein condensate (BEC) of $1.3\!\times\!10^5$ Cs atoms~\cite{Kraemer2004} prepared in the lowest magnetic hyperfine state $\vert F,m_F \rangle=\vert 3,3 \rangle \!\equiv\!|\!\uparrow \rangle$, held in a crossed-beam dipole trap and levitated against gravity by a magnetic field gradient. The BEC is in the Thomas-Fermi regime with the 3D $s$-wave scattering length $a_{\uparrow \uparrow}$ tuned to $a_{\uparrow \uparrow}\!\approx\!220$ $a_0$ corresponding to an offset magnetic field of $B\!=\!21.24(1)$ G. A 2D optical lattice, generated by two retro-reflected laser beams propagating in orthogonal directions, is gradually ramped up in $500$ ms to a potential depth of $30\Er$, with $\Er\!=\! \pi^2\hbar^2/(2ma^2)$ the photon recoil energy, cutting the 3D system into an array of 1D tubes that are oriented along the vertical direction, as sketched in Fig.~\ref{fig1}d. Here, $a\!=\!\lambda/2$ is the lattice spacing with $\lambda\!=\!1064.5$ nm the wavelength of the lattice light. The longitudinal trapping frequency in the 1D tubes is $25.6(3)$ Hz. The magnetic field is then ramped up adiabatically to $B\!=\!35.1$ G, tuning $a_{\uparrow \uparrow}$ to $a_{\uparrow \uparrow}\!\approx\!750$ $a_0$, setting the Lieb-Linger (LL) interaction parameter $\gamma_{\uparrow \uparrow}\!=\!mg_{\uparrow}/(\hbar^2\rho)\sim14$, where $\rho\!=\!N/L\approx\SI{1.33}{\micron}^{-1}$ is the average 1D density and $L$ is the average system length. Here $g_{\uparrow}\approx2\hbar\omega_{\perp}a_{\uparrow\uparrow}$~\cite{meinert2017bloch} and $\omega_{\perp}$ is the transversal trap frequency. For these values, the 1D systems are deeply in the fermionized TG regime \cite{kinoshita2004,haller2009}.

The impurities are Cs atoms that have been transferred to the Zeeman substate $|3,2\rangle\!\equiv\!|\!\downarrow \rangle$ by means of a short radio-frequency pulse. Power and duration are set such that on average one impurity per tube is created. The pulse duration (15 $\mu$s) is much shorter than the Fermi time ($t_\text{F}\!=\!120$ $\mu$s), ensuring that the spatial profile of the impurity closely matches the one of the host gas. The 3D scattering length between impurity and host atoms $a_{\uparrow \downarrow}$ also varies with $B$, see Fig.~\ref{fig1}e. At $B\!=\!35.1$ G the host-impurity LL parameter $\gamma_{\uparrow \downarrow}$ takes the value $\gamma_{\uparrow \downarrow}\sim9$. The impurity atoms in $|\!\downarrow \rangle$ experience a smaller levitating force and would be accelerated by $F_{\downarrow}\!=\!mg/3$. Such a comparatively strong force would lead to a non-adiabatic time evolution~\cite{meinert2017bloch}, populating the continuous spectrum of the gapless quantum liquid and pulling the system away from its ground state (see below). To avoid this, we adiabatically turn on optical levitation in 100 ms. Specifically, a $1064$-nm Gaussian beam with a $1/e^2$ waist of $\sigma_{z}\!\approx\!210$ $\micron$, positioned $\sigma_{z}/2$ above the atoms, generates a nearly linear optical potential gradient. Approximately $10$ W of laser power indiscriminately levitate the host and the impurity atoms when the magnetic force is off. A tunable force $F_{\downarrow}$ on the impurity atoms, while still fully levitating the host atoms, can then be generated by adjusting the fraction of optical vs. magnetic levitation.

\subsection{Role of finite force and finite interaction}
\begin{figure}[t!]
\renewcommand{\figurename}{Fig.}
	\includegraphics[width=\linewidth]{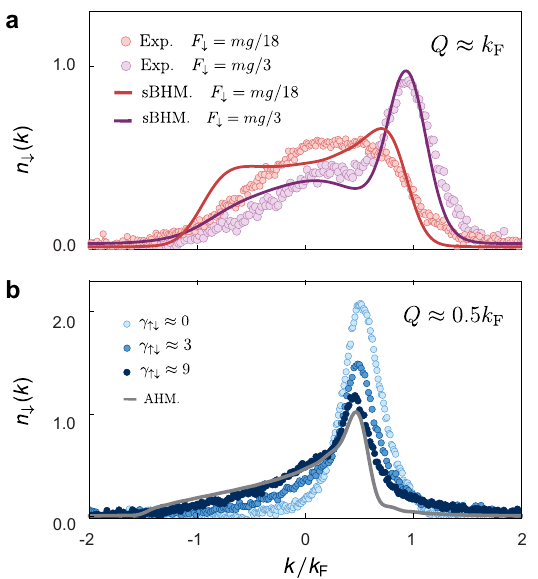}
    \caption{\textbf{Role of finite force and finite interaction.} {\bf{a}} Measured $n_\downarrow(k)$ at fixed $Q$ and fixed  $\gamma_{\uparrow\downarrow}$ for two different values of the force $F_{\downarrow}$ as indicated. Each distribution is the average of $7$ experimental realizations. The experimental data is compared to the results of the simulations based on the sBHM.  {\bf{b}} Measured $n_\downarrow(k)$ at fixed $Q$ for different $\gamma_{\uparrow\downarrow}$. The solid line is the prediction from AHM.
}
\label{extendedFig1}
\end{figure}
Here we study the role of the finite force and finite interaction in our system. In Fig.~\ref{extendedFig1}a, we show $n_\downarrow(k)$ at a fixed total momentum $\hbar Q\approx \hbar k_\text{F}$ for two different values of force $F_{\downarrow}$. For strong force $F_{\downarrow}\!=\!mg/3$, the distribution $n_\downarrow(k)$ is skewed and has a peak around $k=k_\text{F}$. In contrast, for a relatively small force $F_{\downarrow}=mg/18$, the distribution is more symmetric and flat-top as expected for a fermionic distribution. Simulations based on sBHM are in good agreement with our experimental data. The residual asymmetry in the theoretical curve is attributed to the finite $F_{\downarrow}$. The deviation between theory and experiment mainly results from inhomogeneities of the experimental system, in view of the distribution of $k_F$ values for different tubes. We next turn to the effect of finite interaction strength on the momentum distribution. In Fig.~\ref{extendedFig1}b. we show  $n_\downarrow(k)$ at a fixed $Q\approx 0.5k_\text{F}$ for three different values of interaction strength $\gamma_{\uparrow\downarrow}$. Close to the non-interacting point $\gamma_{\uparrow\downarrow}\approx0$, the distribution resembles a bosonic distribution peaked around $k=0.5k_\text{F}$ and does not show any skewness. As we increase the interaction strength to a moderate value of $\gamma_{\uparrow\downarrow}\approx3$, the height of the peak decreases, and $n_\downarrow(k)$ broadens to the left. Only for sufficiently strong interaction, the distribution starts to agree with the prediction from AHM. This confirms that strong interactions are crucial for the emergence of anyonic correlations in our system. Note that the peak in the measured $n_\downarrow(k)$ is broader than the AHM predictions for a single tube. This we again attribute to the effect of inhomogeneities.             

\subsection{Exchange symmetry engineering}
We now elaborate on the way in which the emergence of a spin wave in the system leads to the appearance of anyonic correlations on the original particles as expressed by Eq. (\ref{momentumdistribution}). Due to the phenomenon of spin-charge separation, the exchange symmetry of the spatial part is dictated by the exchange symmetry of the spin part of the wavefunction. To obtain an exchange phase of $\theta$ in the spatial wavefunction, we need to have an exchange phase of $-\theta$ on the spin wavefunction. To describe the system, we will use the bosonic version of the approach described in ref.\cite{PhysRevA.109.012209}, where the spinful bosonic system is replaced by a spinless bosonic charge sector and a spin chain, describing the spin of each atom. The unitary pairwise spin-exchange operators $\hat{\mathcal{E}}_{\ell,\ell'}$ exchange spin $\ell$ with spin $\ell'$ in the spin chain. The set of $\hat{\mathcal{E}}$ operators generates the symmetric group of permutations $S_N$. A fully anyonic wavefunction should be a simultaneous eigenstate of all $\hat{\mathcal{E}}_{\ell,\ell'}$, with the eigenvalue $e^{-i\theta\sign(\ell-\ell')}$. 

This state cannot exist for various reasons. Since $\hat{\mathcal{E}}^2$ is the identity operator, the eigenvalues of $\hat{\mathcal{E}}$ are $\pm 1$, corresponding to triplet (bosonic) and singlet (fermionic) wavefunctions. Furthermore, two exchange operators of the type $\hat{\mathcal{E}}_{\ell,\ell'}$ and $\hat{\mathcal{E}}_{\ell',\ell''}$ do not commute with each other, as can easily be verified.
Simultaneous eigenstates of all pairwise exchange operators are therefore not easy to find, as a result of the fact that the group $S_N$ for $N$ larger than two is non-abelian. Nevertheless, certain observables in the form of correlation functions can be sensitive to only a subgroup of exchanges, as we will show in the following. 

We will now try to find common eigenstates of just a subgroup of $S_N$, with the required form of eigenvalues. In this sense, even though this method cannot generate a fully anyonic wavefunction of the host-impurity system, it can at least give us direct access to specific observables of the anyonic gas. We are looking for a subgroup of $S_N$ whose elements can have complex eigenvalues. Cyclic subgroups are abelian and the eigenvalues of the different elements are given by the $m$-th roots of unity, if $m$ is the size of the cycle. We will concentrate on the cyclic group of maximal order, $C_N$, as this is the most relevant for us. The group generator $\hat{C}$ performs a cyclic rotation of the spin chain configuration of the system $\hat{C}|\sigma_1,..,\sigma_N\rangle=|\sigma_N,\sigma_1,..,\sigma_{N-1}\rangle$. The eigenvalues are given by $e^{-i\theta}$, for $\theta=2\pi n/N$, with $n=0,..,N-1$ and the eigenstates are spin waves. Let us clarify the connection between the exchange phase and the eigenvalue of $\hat{C}$. One cyclic permutation corresponds to $N-1$ backwards binary exchanges. This can be seen by inspecting the effect of the operator on the state of the spin chain. To reproduce the behavior of anyons with forward exchange phase $-\theta$, the eigenvalue of $\hat{C}$ should correspond to $e^{i\theta(N-1)}$. This reduces to $e^{-i\theta}$, using the condition $\theta N= 2\pi n$, with $n \in \mathbb{Z}$, necessary to keep the wavefunction single-valued.
For the case of a single impurity, the spin waves take the form
\begin{equation}
    |\theta\rangle=\frac{1}{\sqrt{N}}\sum_{\ell=0}^{N-1} e^{i\theta \ell}\hat{C}^\ell |\downarrow,\uparrow,\uparrow..\uparrow\rangle.
    \label{spinwave}
\end{equation} 
  
We want to identify correlation functions that are well described by the $\hat{C}$ operator. The simplest example of these is the one-body correlation function of the impurity, for the single-impurity case. To see this connection, consider the action of the operator $\hat{b}_{\downarrow}^\dagger(x)\hat{b}_{\downarrow}(y)$ on the spin configuration of the 1D system. The destruction operator is only non-zero if the spin down particle is found at position $y$, the creation operator then places it at position $x$. As a result, the spin configuration of the system has been shifted by exactly the amount $\hat{N}(x)-\hat{N}(y)$, taking $x>y$. Here $\hat{N}(x)=\int_{-\infty}^x \hat{n}(y)dy$ counts the number of particles to the left of $x$. This corresponds to the application of the operator $\hat{C}^{\hat{N}(x)-\hat{N}(y)}$. We can therefore rewrite
\begin{equation}
\hat{b}_{\downarrow}^\dagger(x)\hat{b}_{\downarrow}(y)=\hat{b}^\dagger(x)\hat{b}(y)  \hat{C}^{\hat{N}(x)-\hat{N}(y)} \hat{\Pi}_\downarrow(\hat{N}(y)),
\end{equation}
where $\hat{b}$ is the destruction operator of spinless hardcore bosons in the charge sector and $\hat{\Pi}_\downarrow(\hat{N}(y))$ is the projector operator on spin down for the spin at position $\hat{N}(y)$ in the spin chain. If the spin state $|\theta\rangle$ is prepared, we get
\begin{align}
\langle\theta|\hat{b}_{\downarrow}^\dagger(x)\hat{b}_{\downarrow}(y)|\theta\rangle=&\frac{1}{N}e^{-i\theta \hat{N} (x)}\hat{b}^\dagger(x)\hat{b}(y)e^{i\theta \hat{N}(y)}= \nonumber \\ &
=\frac{1}{N}\hat{a}^\dagger(x)\hat{a}(y),
\end{align}
where in the last equivalence we used the Jordan-Wigner transformation $\hat{a}=\hat{b}e^{i\theta \hat{N}}$. The factor $1/N$ results from the mean value of $\hat{\Pi}_\downarrow$ on the spin wave.   
It is easy to see how this argument can be generalized to the multi-impurity case, giving a family of anyonic correlation functions that can be exactly simulated with this method. Their explicit expression is given by
\begin{align}
     &\hat{b}^\dagger_{\sigma}(x_1)...\hat{b}^\dagger_{\sigma}(x_m)\hat{b}_{\sigma}(x_1+d)...\hat{b}_{\sigma}(x_m+d)\propto \nonumber\\ &
     \hat{a}^\dagger(x_1)...\hat{a}^\dagger(x_m)\hat{a}(x_1+d)...\hat{a}(x_m+d),
\end{align}
where the number $m$ of creation (destruction) operators should match the number of spin $\sigma$ particles in the spin wave. This demonstrates how, whenever the spin-wave state is realized, we can find correlation functions of the original spinful gas that map exactly onto correlation functions of a system of $N$ anyons, explaining why it is possible to access the momentum distribution of the anyons with local measurements on the original spinful bosons. Making use of this equivalence in practice requires the control of the spin state of the system, but it is completely independent of the state in the charge sector. It is therefore possible to directly measure dynamics of the anyonic correlation functions, assuming the spin wavefunction remains in a spin-wave state during the evolution. In our system, we prepare a spin wave as the eigenstate of momentum with lowest energy by slowly accelerating the impurity.

\subsection{Emergence of anyons via spin-charge separation}
We now turn to a lattice model in order to understand the emergence of anyons in a gas of spinful hardcore bosons prepared in a finite-momentum ground state. We consider the Hamiltonian $\hat{H}_\text{lat}$ describing a gas of $N$ spinful hardcore bosons, 
\begin{equation}
    \hat{H}_\text{lat}=-J\sum_{\ell=1,\sigma}^{L-1} \hat{b}_{ \sigma \ell}^\dagger\hat{b}_{\sigma \ell+1}-J\sum_{\sigma}\hat{b}_{ \sigma L}^\dagger\hat{b}_{\sigma 1}+h.c.
    \label{h_lat}
\end{equation}
Here, $\hat{b}_{\sigma \ell}^\dagger$ ($\hat{b}_{\sigma \ell}$) are bosonic creation (annihilation) operator at site $\ell$ and $\sigma\!=\!(\uparrow,\downarrow)$ is the spin index and $J$ is the hopping amplitude, whose value will be specified below. We assume to be in the low density limit $N/L\ll1$, where $L$ is the number of lattice sites, and we impose periodic boundary conditions so that conservation of momentum is assured. The operators $\hat{b}_{\sigma \ell}^\dagger$ ($\hat{b}_{ \sigma \ell}$) are assumed to satisfy a no-double-occupancy (NDO) constraint, $\sum_{\sigma}\hat{b}_{\sigma \ell}^\dagger\hat{b}_{ \sigma \ell}\leq1$. Under this NDO constraint, spin and charge degrees of freedom separate, $i.e.$, the wavefunction $|\Psi\rangle$ can be written as $|\Psi\rangle\!=\!|\varphi\rangle\otimes|\chi\rangle$. Here, $|\varphi\rangle$ and $|\chi\rangle$ denote the wavefunction for the charge and the spin part, respectively. The Hamiltonian $\hat{H}_\text{lat}$ can be written in spin-charge separated form as~\cite{Kesharpu} 
\begin{equation}
    \hat{H}_\text{sc}=-J\sum_{\ell=1}^{L-1} \hat{f}_{\ell}^\dagger \hat{f}_{\ell+1}-J(-1)^{N-1}\hat{f}_L^\dagger \hat{f}_1\hat{C}^\dagger+h.c.,
\end{equation}

where $\hat{f}_{\ell}^\dagger$ ($\hat{f}_{\ell}$) is the spinless fermionic creation (annihilation) operator at site $j$ and $\hat{C}$ is the spin permutation operator acting on the spin chain as $\hat{C}|\sigma_1,\sigma_2,..,\sigma_{N}\rangle\!=\!|\sigma_{N},\sigma_1,..,\sigma_{N-1}\rangle$. Note that a bosonic description of the charge sector, with hardcore constraint, is also possible, but has the disadvantage that the bosonic particles are still interacting so that diagonalization is not straightforward. The spin permutation operator $\hat{C}$ and the spinless fermionic operators can be diagonalized separately as they are independent of each other. The eigenstates of $\hat{C}$ are spin waves of the form
\begin{equation}
    |\psi_{\nu}\rangle = \frac{1}{\sqrt{N_{\nu}}}\sum_{j=0}^{N_\nu-1} e^{i\theta j}\hat{C}^j |\sigma_1,..,\sigma_N\rangle,
\end{equation}
where $|\sigma_1,..,\sigma_N\rangle$ is an arbitrary configuration of the spin chain, $\nu$ enumerates all possible disconnected spin blocks and $N_\nu$ corresponds to the total number of distinct elements of the form $\hat{C}^j |\sigma_1,...,\sigma_N\rangle$ in the $\nu$-th block. The eigenvalues of $\hat{C}$ are given by $e^{-i\theta}$, for $\theta=2\pi n/N_\nu$, with $n=0,..,N_\nu-1$. In the case of a single impurity $N_{\downarrow}=1$, the eigenstates take the form of Eq.\!\ (\ref{spinwave}). By projecting $\hat{H}_\text{sc}$ on the eigenspace of $\hat{C}$, we get an effective Hamiltonian for the charge sector
\begin{equation}
    \hat{H}_\text{eff}=-J\sum_{\ell=1}^{L-1} \hat{f}^\dagger_{\ell} \hat{f}_{\ell+1}-J(-1)^{N-1}e^{i\theta}\hat{f}^\dagger_{L} \hat{f}_{1}+h.c.
    \label{h_eff}
\end{equation}
Here we see that the fermionic charge sector acquires an overall flux. This spin-generated flux is a collective effect, imposed by the spin waves onto the charge degrees of freedom. Note that the original Hamiltonian $\hat{H}_{\text{lat}}$ does not break time-reversal symmetry (TRS). However, TRS is broken for the $\hat{H}_\text{eff}$ governing the charge sector. Finally, we perform an anyonic transformation
\begin{equation}
   \hat{a}_\ell=\hat{f}_\ell e^{i (\theta+\pi) \hat{N}_\ell} \qquad \text{with} \quad \hat{N}_{\ell}=\sum_{j=1}^{\ell-1}\hat{n}_{j}
   \label{anyontransformation}
\end{equation}
The phase factor in the boundary term vanishes, $(-1)^{N-1}e^{i\theta} e^{i(\theta+\pi) (N-1)}=1$, and the Hamiltonian $\hat{H}_\text{eff}$ can be mapped onto a system of hardcore anyons with a periodic boundary condition
\begin{equation}
    \hat{H}_\text{AHM}=-J\sum_{\ell=1}^{L-1} \hat{a}_{\ell}^\dagger \hat{a}_{\ell+1}-J\hat{a}_L^\dagger \hat{a}_1+h.c.
\end{equation}
As one can see, the anyonic model does not contain any concatenated flux. The transformation Eq.(\ref{anyontransformation}) is a generalized Jordan-Wigner transformation and the anyons can be understood as composite particles in the charge sector~\cite{celi2023,valentírojas2024fluxattachment}. Each spin wave selects a specific value for the statistical phase. In the thermodynamic limit, this result also holds for any choice of boundary conditions. This justifies the use of fixed boundary conditions in the numerics. 

Next, we turn to anyonic observables that can be measured experimentally. The real-space density of these anyons can be extracted by measuring the total density of the gas $\langle \varphi|\hat{a}_{\ell}^\dagger \hat{a}_{\ell}|\varphi\rangle=\langle \varphi|\hat{f}_{\ell}^\dagger \hat{f}_{\ell}| \varphi\rangle=\sum_{\sigma}\langle \Psi|\hat{b}_{\sigma \ell}^{\dagger}\hat{b}_{\sigma \ell}|\Psi\rangle$, where $\Psi$ is the many-body wavefunction of the whole system. However, for hardcore anyons the real-space density is independent of $\theta$. The one-body correlator $\langle\hat{a}_{i}^\dagger \hat{a}_{j}\rangle$, on the other hand, is very sensitive to $\theta$. The Fourier transform of this gives the anyonic momentum distribution, which can be measured by measuring the momentum distribution of the impurity in our system via equation (\ref{momentumdistribution}). Note that Hamiltonian (\ref{h_eff}) can be diagonalized exactly~\cite{Kesharpu}. The momenta of the fermions correspond to the rapidities of the system.

\subsection{Anyon-Hubbard model}

To benchmark the anyonic behavior realized in the experiment, we next elaborate on the anyonic correlations of the paradigmatic AHM, which can be effectively simulated by using a bosonic model with density-dependent tunneling.
By using a fractional version of the Jordan-Wigner transformation, i.e., the anyon-boson mapping
\begin{equation}
    \hat{a}_{\ell}=\hat{b}_{\ell}e^{i\theta \hat{N}_{\ell}},~~~\hat{N}_{\ell}=\sum_{j=1}^{\ell-1}\hat{n}_{j},
    \label{AB_map}
\end{equation}
 
the anyon-Hubbard model from Eq.\ (\ref{AHM}) can be expressed in terms of bosonic operators as
\begin{equation}
    \hat{H}_{\text{AHM}}^\text{B}=-J\sum_{\ell}\left(\hat{b}_{\ell}^{\dagger}\hat{b}_{\ell+1}e^{i\theta\hat{n}_{\ell}}+h.c.\right)+\frac{U}{2}\sum_{\ell}\hat{n}_{\ell}\left(\hat{n}_{\ell}-1\right).
    \label{H_AM_B}
\end{equation}

Here, $\hat{b}_\ell$ are the bosonic annihilation operators at site $\ell$.

Different from the the bosonic one-body density correlation $\langle\hat{b}_{\ell}^{\dagger}\hat{b}_{\ell'}\rangle$, the correlator of anyons $\langle\hat{a}_{\ell}^{\dagger}\hat{a}_{\ell'}\rangle$ can be expressed as
\begin{align}
    \langle\hat{a}_{\ell}^{\dagger}\hat{a}_{\ell'}\rangle = \langle\hat{b}_{\ell}^{\dagger}e^{i\theta(\hat{N}_{\ell'}-\hat{N}_{\ell})}\hat{b}_{\ell'}\rangle.
    \label{corr_ab_AHM}
\end{align}

For the data shown in Fig.~\ref{fig2} and Fig.~\ref{Fig3}, we have assumed $N\!=\!10$ anyons in $L=40$ lattice sites in the hard-core limit. 

\subsection{Dynamical evolution with sBHM}
In practice, a spin wave can be generated by slowly accelerating the impurity. In order to efficiently simulate such a dynamical process, we consider a spinful Bose-Hubbard model (sBHM) on a 1D lattice,
\begin{align}
    \hat{H}_{\text{sBHM}}= & -J\sum_{\ell=1}^{L-1}\left(\hat{b}_{\text{\ensuremath{\uparrow}}\ell}^{\dagger}\hat{b}_{\ensuremath{\uparrow}\ell+1}+\hat{b}_{\downarrow\ell}^{\dagger}\hat{b}_{\downarrow\ell+1}+h.c.\right)
    \nonumber \\ &
    +U_{\uparrow\downarrow}\sum_{\ell}\hat{n}_{\uparrow\ell}\hat{n}_{\downarrow\ell}+\sum_{\ell}Fa\ell\hat{n}_{\downarrow\ell}.
    \label{H_BHM}
\end{align}
Here, $\hat{b}_{\uparrow\ell}$ and $\hat{b}_{\downarrow\ell}$ are the annihilation operators of host particles and an impurity at site $\ell$, respectively, with their hopping strength being denoted by $J$. We consider the hardcore limit of the intra-component interaction, i.e., $U_{\uparrow\uparrow}\rightarrow\infty$ and $U_{\downarrow\downarrow}\rightarrow\infty$. The on-site interaction between host particles and the impurity is denoted by $U_{\uparrow\downarrow}$. A constant force $F_{\downarrow}$ is applied only to the impurity. 

We define the dimensionless force $\mathcal{F}=\frac{F_{\downarrow} m}{\hbar^{2}\rho^{3}}$. In the low filling limit, any lattice model reduces to a continuum model with the effective mass given by
\begin{equation}
    m=\frac{\hbar^{2}}{2Ja^{2}}.
\end{equation}
By setting the value of the effective mass to be equal to the particle's mass, we fix the value of $J$.
By defining the filling factor in a lattice $n\!=\!N/L_\text{S}$ with $L_\text{S}$ being the number of lattice sites and $a\!=\!L/L_\text{S}$ being the lattice constant, one obtains the following mapping between quantities,
\begin{align}
    \frac{U}{J}= & \frac{g_{\uparrow\downarrow}}{a}\frac{2ma^{2}}{\hbar^{2}}=2\gamma_{\uparrow\downarrow}\frac{N}{L_\text{S}},
    \\
    \frac{F_{\downarrow}a}{J}= & F_{\downarrow}a\frac{2ma^{2}}{\hbar^{2}}=2\mathcal{F}\left(\frac{N}{L_\text{S}}\right)^{3}.
\end{align}
In our simulation, the initial impurity distribution is defined by the ground state of the Hamiltonian (\ref{H_BHM}) with $F_{\downarrow}\!=\!0$ and a harmonic trapping potential $V$ applied only for the impurity. At $t\!=\!0$, we suddenly remove the traps and switch on the constant force $F_{\downarrow}$. We simulate the quench dynamics by solving the time-dependent Schr\"odinger equation associated with the Hamiltonian (\ref{H_BHM}) by using the time-dependent variational principle (TDVP) based on matrix product states implemented using ITensors~\cite{2022_itensor,2022_itensor03}. The results are presented in Fig.~\ref{fig2} and Fig.~\ref{Fig3}. The parameters have been chosen as $L\!=\!40, N_\downarrow\!=\!1, N_\uparrow\!=\!20, U/J\!=\!9.1$ and $F_{\downarrow}a/J\!=\!0.15$ for numerical convenience. A more costly simulation by using a larger system size (e.g.,\ $L\!=\!120$) at lower filling (e.g.\ $N_\uparrow/L\!=\!0.25$) gives very similar results~\cite{SM}.

\subsection{Swap model}
 
Inspired by the central role of the spin wave in the emergence of anyonic behavior of our system, we develop a toy model whose ground state encodes the spin wave we are targeting,
\begin{align}
\hat{H}_{\text{swap}}= & -J\sum_{\ell=1}^{L-1}\hat{b}_{\text{\ensuremath{\uparrow}}\ell}^{\dagger}\hat{b}_{\ensuremath{\uparrow}\ell+1}-J\sum_{\ell=1}^{L-1}\hat{b}_{\downarrow\ell}^{\dagger}\hat{b}_{\downarrow\ell+1}
\nonumber \\ &
-J_{\text{ex}}e^{-i\theta}\sum_{\ell=1}^{L-1}\hat{b}_{\uparrow\ell}^{\dagger}\hat{b}_{\downarrow\ell+1}^{\dagger}\hat{b}_{\downarrow\ell}\hat{b}_{\uparrow\ell+1}+h.c.,
   \label{H_Jex}
\end{align}
with $\hat{b}_{\uparrow\ell}$ and $\hat{b}_{\downarrow\ell}$ being the annihilation operators of host particles and the impurity at site $\ell$, respectively, and their hopping strength denoted by $J$. In the strongly interacting regime, the swapping strength $J_{\text{ex}}$ is expected to be of the order of $J^2/U_{\uparrow\downarrow}$. 
Importantly, we encode the spin wave information by assigning the factor $e^{-i\theta}$ to the swapping terms. The ground states of the swap model is expected to effectively describe the adiabatic time evolution of generating spin-wave states~\cite{Swap}. 

In a spin-charge separated representation, the one-body correlation function $\langle\hat{b}_{\downarrow\ell}^{\dagger}\hat{b}_{\downarrow\ell'}\rangle$ of the single impurity can be implemented by hopping of spinless particles and swapping $\hat{\mathcal{E}}$ on the spin chain~\cite{PhysRevA.108.063315,PhysRevA.109.012209}. Taking $\ell'\geqslant\ell$ as an example, we have
\begin{align}
    \langle\hat{b}_{\downarrow\ell}^{\dagger}\hat{b}_{\downarrow\ell'}\rangle = &      \sum_{m',m}\langle\varphi|\hat{b}_{\ell}^{\dagger}\hat{b}_{\ell'}\delta_{m,\hat{N}_l}\delta_{m',\hat{N}_{l'}}|\varphi\rangle
    \nonumber \\ & \times\langle\chi|\hat{\mathcal{E}}_{m,m+1}\cdots\hat{\mathcal{E}}_{m'-1,m'}|\chi\rangle,
    \label{swap_corr_1}   
\end{align}
Here, the Kronecker $\delta$ operators ensure that the $\ell'$-th site is occupied the $m'$-th spin, and after hopping, the $\ell$-th site is occupied by the $m$-th spin. In the case of a single impurity, the product of swap operators is related to the $\hat{C}$ operator as shown previously, giving rise to a spin wave, which leads to the following one-body correlator of the impurity~\cite{Swap},
\begin{equation}
    \langle\hat{b}_{\downarrow\ell}^{\dagger}\hat{b}_{\downarrow\ell'}\rangle =  \frac{1}{N}\langle\varphi|\hat{b}_{\ell}^{\dagger}\hat{b}_{\ell'} e^{i\theta(\hat{N}_{l'}-\hat{N}_l)}|\varphi\rangle.
\end{equation}
The above equation establishes a direct link between the one-body correlator of the impurity and anyonic correlator Eq.~(\ref{corr_ab_AHM}) and thus Eq.~(\ref{momentumdistribution}) is recovered. Based on a Fourier transformation and using the parameters $L=120, N_\downarrow=1, N_\uparrow=30$ and $J_\text{ex}/J=0.01$, we obtain the quasi-momentum distribution of the impurity shown in Fig.~\ref{fig2} and Fig.~\ref{Fig3}. Note that the small value of swapping strength $J_\text{ex}$ is related to the strong host-impurity interaction and the agreement with experimental data is found for a wide parameter regime~\cite{SM}.

\subsection{Rapidity of anyons in 1D}

In Fig.~\ref{fig4}(d-f), we present the results of the simulation of the quench dynamics of anyonic gases after suddenly removing the harmonic trap in 1D. 
The momentum distribution as a function of evolution time is expressed as
\begin{equation}
    n_\text{a}(k,t)=\frac{1}{2\pi}\iint\text{d}x\text{d}ye^{ik(x-y)}\rho_{\text{HCA}}(x,y;t),
\end{equation}
with the single-particle density matrix of hardcore anyons $\rho_{\text{HCA}}(x,y;t)$.
Following Ref.~\cite{delcampo}, it can be efficiently computed as
\begin{equation}
    \rho_{\text{HCA}}(x,y;t)={\displaystyle \sum_{m,n=0}^{N-1}\phi_{m}^{*}(x,t)A_{mn}(x,y;t)\phi_{n}(y,t)},
\end{equation}
where $A_{mn}(x,y;t)$ are the matrix elements of $\mathbf{A}(x,y;t)=(\mathbf{P}^{-1})^{T}\text{det}\mathbf{P}$, and the elements of matrix $\mathbf{P}(x,y;t)$ are
$P_{mn}(x,y;t)=\delta_{mn}-\left(1-e^{-i\theta\textnormal{sgn}(y-x)}\right)\textnormal{sgn}(y-x)\int_{x}^{y}dz\phi_{m}^{*}(z,t)\phi_{n}(z,t)$.
Here, $\phi_{n}(x,0)$ are the single-particle wavefunctions of the 1D harmonic oscillator, and $\phi_{n}(x,t)$ fulfill the time-dependent Schr\"odinger equation
\begin{equation}
    i\hbar\frac{\partial\phi_{n}(x,t)}{\partial t}=\left(-\frac{\hbar^{2}}{2m}\frac{\partial^{2}}{\partial x^{2}}+\frac{m\omega_0^{2}x^{2}\Theta(-t)}{2}\right)\phi_{n}(x,t),
\end{equation}
with Heaviside step function $\Theta(t)$, which models a sudden quench $\omega(t)=\omega_{0}\Theta(-t)$. The solution is found to be $\phi_{n}(x,t)=\phi_{n}(x/b(t),0)e^{imx^{2}\dot{b}/2b\hbar-iE_{n}\tau(t)/\hbar}/\sqrt{b(t)}$, with the scaling factor $b(t)=\sqrt{1+\omega_{0}^{2}t^{2}}$, $\tau(t)=\int_{0}^{t}dt'/b^{2}(t')$ and $E_{n}=\hbar\omega_{0}(n+1/2)$. In the experiment, the trapping frequency is set to $\omega_0=25.6(3)$ Hz, and the average Fermi time is $t_{\rm F}=2m/\hbar k_{\rm F}^2\approx 0.12$ ms.

\cleardoublepage

\beginsupplement

\begin{onecolumngrid}
\begin{center}
{\bf {\large Supplementary Materials of \\ ``Anyonization of bosons''}}\\
\end{center}
\end{onecolumngrid}

\maketitle

\section{Exact solution via Bethe ansatz}

Here, we consider the problem of an impurity interacting with a one-dimensional TG gas via a short-ranged $\delta$-function potential of arbitrary strength $g_{\uparrow\downarrow}$. The TG gas can be mapped to a gas of spin-polarized free fermions. The system is governed by the Hamiltonian~\cite{Mathy2012}
\begin{equation}\label{eqn1s}
\hat{H} = \frac{1}{2m} \sum_{i=1}^N \hat{P}_i^2 + \frac{1}{2m} \hat{P}_\downarrow^2 + g_{\uparrow\downarrow}\sum_{i=1}^N \delta(x_i-x_\downarrow),
\end{equation}
where $x_{i}$ and $\hat{P}_{i}$ are the position and momentum of the $i$-th background particle, respectively. The position and momentum of the impurity are denoted by $x_\downarrow$ and $\hat{P}_\downarrow$, respectively. All particles are assumed to have the same mass $m$. The dimensionless LL interaction strength is given by $\gamma_{\uparrow\downarrow}\!=\!\frac{mg_{\uparrow\downarrow}}{\hbar^2\rho}$, where $\rho\!=\!\frac{N}{L}$ is the density of the background gas and $L$ is the system-size. The Fermi momentum, defined as $k_\text{F}\!=\!\pi\rho$, is directly proportional to the 1D density $\rho$. This model (\ref{eqn1s}) is integrable for any value of $g_{\uparrow\downarrow}$ and can be solved via Bethe ansatz~\cite{McGuire1965}. Some calculations are easier in the mobile impurity reference frame. This is done via Lee-Low-Pines (LLP) transformation~\cite{LLP1953}, sometimes called the polaron transformation, which is frequently used in polaron physics. For simplicity, we set $g_{\uparrow\downarrow}\equiv g$ and $\hbar=m=1$ in the following.
\subsection{Lee-Low-Pines transformation}
The key object is the operator 
\begin{equation}
\mathcal{Q} = e^{i\hat{P}_\uparrow \hat{x}_\downarrow}.
\end{equation}
Here, $x_\downarrow$ is the position of the impurity and $P_\uparrow$ is the total momentum of the host particles. The transformation of an arbitrary operator $\mathcal{O}$ from the laboratory to the mobile impurity reference frame is given by
\begin{equation}
\mathcal{O} \to \mathcal{O}_{\mathcal{Q}} = \mathcal{Q} \mathcal{O} \mathcal{Q}^{-1}.
\end{equation}
The LLP transformation does not affect the momentum of the host particles but changes the momentum operator of the impurity
\begin{equation}
\hat{P}_{\uparrow\mathcal{Q}} = \hat{P}_\uparrow, \qquad \hat{P}_{\downarrow\mathcal{Q}} = \hat{P}_\downarrow - \hat{P}_\uparrow.
\end{equation}
Therefore, the total momentum of the system in the mobile impurity reference frame reads
\begin{equation}
\hat{P}_{\mathcal{Q}} = \hat{P}_\downarrow .
\end{equation}
Let us apply $\mathcal{Q}$ to the wavefunction. Recall that
\begin{equation} \label{translation}
e^{a\frac{d}{dx}}f(x) = f(x+a).
\end{equation}
As a result,
\begin{equation}
    \mathcal{Q} \Psi_{Q}(x_\downarrow,x_1,\ldots,x_N)= \Psi_{Q}(x_\downarrow,x_1+x_\downarrow,\ldots,x_N+x_\downarrow) = e^{iQ x_\downarrow} \Psi_{Q}(0,x_1,\ldots,x_N). \label{eq:Qfact}
\end{equation}
Here, the subscript $Q$ indicates the value of the total momentum of the system. Note that we work with fermions in the continuum, hence our system is translationally invariant. The shift of all coordinates is achieved by the action of the momentum operator as follows from Eq.~\eqref{translation}. This is how we got the right hand side of Eq.~\eqref{eq:Qfact}. We can rewrite Eq.~\eqref{eq:Qfact} as
\begin{equation}
\Psi_{Q}(x_\downarrow,x_1,\ldots,x_N) = e^{iQ x_\downarrow} \Psi_{Q}(0,x_1- x_\downarrow ,\ldots,x_N - x_\downarrow)\\ \equiv e^{iQ x_\downarrow} f_Q(y_1,\ldots,y_N),
\end{equation}
where $y_j = x_j - x_\downarrow$, $j=1,\ldots,N$.

The function $f_Q(y_1,\ldots,y_N)$ is the wavefunction of the system in the mobile impurity reference frame. Working with its first-quantized representation, we aim at doing an exact calculation for finite $N$. Hence, we impose periodic boundary conditions to take into account finite-size effects. For the following calculations, we consider the case where $N$ is odd.

\subsection{Bethe ansatz solution for arbitrary coupling}
The Hamiltonian~\ref{eqn1s} in the mobile impurity frame transforms to, 
\begin{equation}
\hat{H}_Q = \frac12 \sum_{i=1}^N \hat{P}_i^2 + \frac12(\hat{P}_\downarrow-\hat{P}_\uparrow)^2 + g\sum_{i=1}^N \delta(y_i).
\end{equation}
Thus, any gas particle in the impurity frame is scattered by the impurity particle positioned at the origin. The wavefunctions of the problem in the impurity frame look particularly simple. They are just Slater determinants
\begin{equation} \label{fdef}
f_Q(y_1,\ldots,y_N) = \frac{Y}{\sqrt{N!L^N}} \left|\begin{matrix} e^{ik_1 y_1}& \ldots & e^{ik_{N+1}y_1} \\ \vdots & \ddots & \vdots \\ e^{ik_1 y_N} & \ldots & e^{ik_{N+1}y_N} \\ 
\nu(k_1) & \ldots & \nu(k_{N+1})\end{matrix}\right|, \qquad 0\le y_j\le L,
\end{equation}
where
\begin{equation} \label{nudef}
\nu(q) = \frac{g}2 \frac1{q-\frac{g}2(\Lambda+i)}.
\end{equation}
The factor $Y$ ensures the normalization condition. The set of quasi-momenta $k_1,\ldots,k_{N+1}$ satisfies a system of nonlinear equations (Bethe equations)
\begin{equation} \label{Bethedef}
\cot\frac{k_j L}2 = \frac{2k_j}{g} - \Lambda, \qquad j=1,2,\ldots,N+1.
\end{equation}
These equations are connected to each other by the condition 
\begin{equation}
Q = \sum_{j=1}^{N+1} k_j.
\end{equation}
That is, the sum of the quasi-momenta give the total momentum (which is an observable). In the systen with finite number of particles the total momentum is quantized as usual,
\begin{equation} \label{momentumtotalquantization}
Q = \frac{2\pi}L n, \qquad n=0,\pm1, \pm 2,\ldots.
\end{equation}
The energy of the state is
\begin{equation}
E_\text{F} = \frac12 \sum_{j=1}^{N+1} k_j^2.
\end{equation}

Let us recall how Eqs.~\eqref{Bethedef} are obtained. The function~\eqref{fdef} has to be continuous,
\begin{equation} \label{boundary1}
f_Q(y_1,\ldots,y_N)\left|_{y_j=0}^{y_j=L}\right.= 0.
\end{equation}
Its first derivative should experience a jump such that the second derivative generates the terms $g\delta(y_j)$:
\begin{equation} \label{boundary2}
-\partial_{y_j} f_Q(y_1,\ldots,y_N) \left|_{y_j=0}^{y_j=L}\right. = g f_Q(y_1,\ldots,y_j=0,\ldots,y_N).
\end{equation}
Substituting the function~\eqref{fdef} into these two equations we get the desired equations~\eqref{Bethedef}.

The first important feature of the function~\eqref{fdef} stemming from the Bethe equations is the non-periodicity of the plane waves, $e^{ik_j L}\ne 1$. It is only the function itself that is periodic. The second important feature is that the form~\eqref{fdef} is valid when all $y_j$ are in the interval from zero to $L$. The expression, for example, for $y_j$ in the interval from $L$ to $2L$ is not given by Eq.~\eqref{fdef}. We discuss how to extend Eq.~\eqref{fdef} from the interval $0\le y_j\le L$ for the particular case $g\to\infty$ in the next section.

\subsection{Bethe ansatz solution in the limit of infinite repulsion}
The form of Eq.~\eqref{fdef} further simplifies in the limit of infinite repulsion, $g\to\infty$. There, the function~\eqref{nudef} becomes momentum-independent,
\begin{equation} \label{nuinfdef}
\nu(q) = -\frac1{\Lambda+i}, \qquad g\to\infty
\end{equation}
and the wavefunction~\eqref{fdef} takes the form
\begin{equation} \label{finfdef}
f_Q(y_1,\ldots,y_N) = \frac{\tilde{Y}}{\sqrt{N!L^N}} \left|\begin{matrix} e^{ik_1 y_1}& \ldots & e^{ik_{N+1}y_1} \\ \vdots & \ddots & \vdots \\ e^{ik_1 y_N} & \ldots & e^{ik_{N+1}y_N} \\ 
1 & \ldots & 1\end{matrix}\right|, \qquad g\to\infty
\end{equation}
in the domain $0\le y_j\le L$. The Bethe equations~\eqref{Bethedef} also simplify a lot:
\begin{equation} \label{Betheinfdef}
\cot\frac{k_j L}2 =  - \Lambda, \qquad j=1,2,\ldots,N+1, \qquad g\to\infty.
\end{equation}
We see that the quasi-momenta $k_j$ are quantized like free fermions plus a shift, same for all $k_j$s from a given set:
\begin{equation} \label{Bethe2infdef}
k_j =  q_j +\frac{\mu}L, \qquad j=1,2,\ldots,N+1, \qquad g\to\infty,
\end{equation}
where $q_j$ are free-fermion momenta
\begin{equation} 
q_j =  \frac{2\pi}L n_j, \qquad n_j=0, \pm1, \pm2,\ldots.
\end{equation}
We therefore have
\begin{equation} \label{finf2def}
f_Q(y_1,\ldots,y_N) = \frac{\tilde{Y}}{\sqrt{N!L^N}} \prod_{j=1}^N e^{i\mu y_j/L}\left|\begin{matrix} e^{iq_1 y_1}& \ldots & e^{iq_{N+1}y_1} \\ \vdots & \ddots & \vdots \\ e^{iq_1 y_N} & \ldots & e^{iq_{N+1}y_N} \\ 
1 & \ldots & 1\end{matrix}\right|, \qquad g\to\infty
\end{equation}
in the domain $0\le y_j\le L$. 

It is worth mentioning that the function $f_Q$ remains far from trivial even in the $g\to \infty$ limit, despite the seemingly ``free-fermion'' form of the expressions~\eqref{finfdef} and~\eqref{finf2def}. This is because each plane wave, $e^{ik_j y}$, still does not satisfy periodic boundary conditions, that is, $e^{ik_j L}\ne 1$.

It is important to keep in mind that the function~\eqref{finfdef} is defined in the domain $0\le y_j\le L$ (which means that the gas particles are positioned to the right of the impurity). Let us now extend the definition to the case where some particles are placed to the left of the impurity, that is, we tackle the domain $-L\le y_j\le L$. For that we use the periodicity of $f_Q$ on a ring of circumference $L$:
\begin{equation} \label{fqperiod}
f_Q(y_1,\ldots,y_j-L,\ldots,y_N) = f_Q(y_1,\ldots,y_j,\ldots,y_N), \quad j=1,\ldots,N.
\end{equation}
Having Eq.~\eqref{finfdef} and using Eq.~\eqref{fqperiod} we get
\begin{equation} \label{finfdeffull}
f_Q(y_1,\ldots,y_N) = \frac{\tilde{Y} e^{-i\frac\mu2 \prod_{j=1}^N \sign(y_j)}}{\sqrt{N!L^N}} \left|\begin{matrix} e^{ik_1 y_1}& \ldots & e^{ik_{N+1}y_1} \\ \vdots & \ddots & \vdots \\ e^{ik_1 y_N} & \ldots & e^{ik_{N+1}y_N} \\ 
1 & \ldots & 1\end{matrix}\right|, \qquad g\to\infty.
\end{equation}
It is the factor containing the sign functions that ensures the validity of Eq.~\eqref{finfdeffull} in the whole domain $-L\le y_j\le L$.
Coming back to the laboratory frame we get for this wavefunction
\begin{equation} \label{finf3def}
\Psi_Q(x_\downarrow, x_1,\ldots,x_N) = \frac{\tilde{Y} e^{-i\frac\mu2 \prod_{j=1}^N \sign(x_j-x_\downarrow)}}{\sqrt{N!L^N}}\\
\times\left|\begin{matrix} e^{ik_1 x_1}& \ldots & e^{ik_{N+1} x_1} \\ \vdots & \ddots & \vdots \\ e^{ik_1 x_N} & \ldots & e^{ik_{N+1} x_N} \\ 
e^{ik_1 x_\downarrow} & \ldots & e^{ik_{N+1}x_\downarrow} \end{matrix}\right|, \qquad g\to\infty.
\end{equation}
valid in the domain $-L/2\le x_j\le L/2$, $j=1,\ldots,N,\downarrow$. Recall that Eqs.~\eqref{finfdeffull} and \eqref{finf3def} are connected by the transformation~\eqref{eq:Qfact}. The function~\eqref{finf3def} is antisymmetric with respect to any permutation of the host particles. 

At this point, let us summarize our knowledge about the wavefunctions and about the spectrum in the $g\to\infty$ limit. Equation~\eqref{finf3def} is a Slater determinant. This way, the impurity problem at infinite repulsion behaves just as the free fermion one. Indeed, if any two of the coordinates from the set $x_1,\ldots,x_N, x_\downarrow$ take the same value, the determinant vanishes regardless of the values of the quasi-momenta $k_1,\ldots,k_{N+1}$. This is how fermions should behave. The function~\eqref{finf3def} is antisymmetric with respect to any permutation of the host particles. What is really amazing about Eq.~\eqref{finf3def} is its periodicity, that is, $\Psi$ takes the same values at $x_j =-L/2 $ and $ x_j =L/2$, $j=1,\ldots,N,\downarrow$ in the case of $k_1,\ldots,k_{N+1}$ quantized according to the Bethe equations~\eqref{Betheinfdef} and~\eqref{momentumtotalquantization}, despite the fact that each plane wave is not periodic in this interval, $e^{ik_jL}=e^{i\mu}$. We now reformulate the problem in the language of the second quantization.
\subsection{Anyon-fermion mapping of the $g\to\infty$ problem and second quantization}

The parameter $\mu$ is related to the total momentum $\hbar Q$ as $Q=k_\text{F}(1-\mu)$. The wavefunction for each value of $\mu$ can be written in the second-quantized form as follows 
\begin{equation} \label{psiqsecond}
|\Psi_Q\rangle = \frac1{\sqrt{N!}} \int_0^L dx_\downarrow dx_1\cdots dx_N \Psi_Q(x_\downarrow,x_1,\ldots,x_N) \psi^\dagger_\text{A}(x_\downarrow) \psi^\dagger(x_1)\cdots \psi^\dagger(x_N)|0\rangle.
\end{equation}
Here, $\psi$ is the fermion destruction operator and $\psi_\text{A}$ behaves as an impenetrable anyon with respect to $\psi$,
\begin{equation} \label{psiapsi}
\psi_\text{A}(x_\downarrow)\psi(x) +e^{-i\pi\mu\sign(x_\downarrow-x)}\psi(x)\psi_\text{A}(x_\downarrow) = 0,
\end{equation}

while the host particles behave with respect to each other as free fermions. Note that the momentum distribution of the impurity $n_{\downarrow}(k)$ is the same whether the host particles are free fermions or a TG gas and $n_{\downarrow}(k)$ can be expressed through a correlation function of 1D impenetrable anyons~\cite{Gamayun2020}.

\section{Anyon-Hubbard model in a nutshell}
In this section, we briefly review the properties of the above-mentioned anyon-Hubbard model, which is a paradigmatic model to describe anyons in 1D lattices~\cite{Keilmann2011,Axel2015},
\begin{equation}
    \hat{H}_{\text{AHM}}=-J\sum_{\ell}\left(\hat{a}_{\ell}^{\dagger}\hat{a}_{\ell+1}+h.c.\right)+\frac{U}{2}\sum_{\ell}\hat{n}_{\ell}\left(\hat{n}_{\ell}-1\right),
    \label{H_AHM}
\end{equation}
where $J$ and $U$ denote the tunneling amplitude and the on-site interaction between anyons, respectively, and $\hat{n}_{\ell}=\hat{a}_{\ell}^{\dagger}\hat{a}_{\ell} $ is the number operator at site $\ell$. The \textit{anyonic} operators $\hat{a}_\ell$ obey the generalized commutation relations
\begin{equation}
    \hat{a}_{j}^{\dagger}\hat{a}_{k}-e^{-i\theta\text{sgn}(j-k)}\hat{a}_{k}^{\dagger}\hat{a}_{j}=\delta_{jk},~~~~\hat{a}_{j}\hat{a}_{k}-e^{-i\theta\text{sgn}(j-k)}\hat{a}_{k}\hat{a}_{j}=0.
\end{equation}
The above relations can be obtained by means of a fractional version of the Jordan-Wigner transformation, i.e., via the anyon-boson mapping
\begin{equation}
    \hat{a}_{\ell}=\hat{b}_{\ell}e^{i\theta N_{\ell}},~~~\hat{N}_{\ell}=\sum_{j=1}^{\ell-1}\hat{n}_{j},
    \label{AB_map}
\end{equation}
where $\hat{b}_\ell$ are bosonic operators and obey the bosonic commutation relation $[\hat{b}_{j},\hat{b}_{k}^{\dagger}]=\delta_{jk},[\hat{b}_{j},\hat{b}_{k}]=0=[\hat{b}_{j}^{\dagger},\hat{b}_{k}^{\dagger}]$. Note that the above transformation gives the same number operators, i.e., $\hat{n}_{\ell}=\hat{a}_{\ell}^{\dagger}\hat{a}_{\ell}=\hat{b}_{\ell}^{\dagger}\hat{b}_{\ell}$.

Combining the anyon-boson mapping (\ref{AB_map}) with Eq.\ (\ref{H_AHM}), the anyon-Hubbard Hamiltonian can be expressed in terms of bosonic operators as
\begin{equation}
    \hat{H}_{\text{AHM}}^\text{B}=-J\sum_{\ell}\left(\hat{b}_{\ell}^{\dagger}\hat{b}_{\ell+1}e^{i\theta\hat{n}_{\ell}}+h.c.\right)+\frac{U}{2}\sum_{\ell}\hat{n}_{\ell}\left(\hat{n}_{\ell}-1\right).
    \label{H_AM_B}
\end{equation}
Here, we will be interested in the quasi-momentum distribution. One has to distinguish the quasi-momentum distribution in terms of bosonic operators from that using anyonic operators,
\begin{align}
    \langle\hat{n}_{k}^\text{b}\rangle= & \frac{1}{L}\sum_{\ell\ell'}e^{ik(\ell-\ell')}\langle\hat{b}_{\ell}^{\dagger}\hat{b}_{\ell'}\rangle,
    \\
    \langle\hat{n}_{k}^\text{a}\rangle= & \frac{1}{L}\sum_{\ell\ell'}e^{ik(\ell-\ell')}\langle\hat{a}_{\ell}^{\dagger}\hat{a}_{\ell'}\rangle.
\end{align}
In hardcore limit, the bosonic Hamiltonian (\ref{H_AM_B}) will be independent of the statistical angle $\theta$, i.e.,
\begin{align}
    \hat{H}_{\text{AHM}}^\text{B} \stackrel{U\rightarrow\infty}{\longrightarrow}\hat{H}_{\text{AHM}}^\text{B}=-J\sum_{\ell}\left(\hat{b}_{\ell}^{\dagger}\hat{b}_{\ell+1}+h.c.\right).
\end{align}
Thus, computing the quasi momentum of bosons $\langle\hat{n}_{k}^\text{b}\rangle$ would give a $\theta$-independent quasi-momentum distribution~\cite{2015Tang}.

However, the anyonic nature can be revealed by computing the anyonic quasi-momentum distribution $\langle\hat{n}_{k}^\text{a}\rangle$, because the anyonic correlation $\langle\hat{a}_{\ell}^{\dagger}\hat{a}_{\ell'}\rangle$ are modified by the Jordan-Wigner transformation. To clarify this, we can explicitly write the anyonic correlation as
\begin{align}
    \langle\hat{a}_{\ell}^{\dagger}\hat{a}_{\ell'}\rangle = \langle\hat{b}_{\ell}^{\dagger}e^{i\theta(\hat{N}_{\ell'}-\hat{N}_{\ell})}\hat{b}_{\ell'}\rangle
    =\begin{cases}
    \langle\hat{b}_{\ell}^{\dagger}\hat{b}_{\ell'}e^{i\theta(\hat{N}_{\ell'}-\hat{N}_{\ell})}\rangle, & \ell'\geq\ell\\
    \langle\hat{b}_{\ell}^{\dagger}\hat{b}_{\ell'}e^{i\theta(\hat{N}_{\ell'}-\hat{N}_{\ell}+1)}\rangle, & \ell'<\ell.
    \end{cases}
    \label{corr_ab}
\end{align}
Here, $\langle\bullet\rangle\equiv\langle\Psi|\bullet|\Psi\rangle$ represents the expectation value with respect to the ground state $|\Psi\rangle$ of the Hamiltonian (\ref{H_AM_B}), which can be expressed in the Fock state basis $\{|\psi_{i}\rangle\}$ as
\begin{equation}
    |\Psi\rangle=\sum_{j}c_{j}|\psi_{j}\rangle,~~~~|\psi_{j}\rangle=|n_{1}^{j},n_{2}^{j},\cdots,n_{L}^{j}\rangle
    \label{AHM_basis}
\end{equation}
with complex coefficients $c_i$, $n_{\ell}^{i}=\{0, 1\}$ in the hard-core limit, and $L$ being the total site number of the chain. 
In this case, the above anyonic correlation (\ref{corr_ab}) can be rewritten as
\begin{equation}
    \langle\hat{a}_{\ell}^{\dagger}\hat{a}_{\ell'}\rangle=\begin{cases}
    {\displaystyle \sum_{i,j}}c_{i}^{*}c_{j}\langle\psi_{i}|\hat{b}_{\ell}^{\dagger}\hat{b}_{\ell'}|\psi_{j}\rangle e^{i\theta(N_{\ell'}^{j}-N_{\ell}^{j})}, & \ell'\geq\ell,\\
    {\displaystyle \sum_{i,j}}c_{i}^{*}c_{j}\langle\psi_{i}|\hat{b}_{\ell}^{\dagger}\hat{b}_{\ell'}|\psi_{j}\rangle e^{-i\theta(N_{\ell}^{j}-N_{\ell'}^{j}-1)}, & \ell'<\ell.
    \end{cases}
    \label{corr_ab_fock}
\end{equation}
with $N_{\ell}^{j}=\sum_j$. Here, the ($N$-dependent) phase factors in Eq.\ (\ref{corr_ab}) and (\ref{corr_ab_fock}) are attributed to the contribution from the Jordan-Wigner transformation, which transfers the bosonic correlation $\langle\hat{b}_{\ell}^{\dagger}\hat{b}_{\ell'}\rangle$ to be the anyonic one $\langle\hat{a}_{\ell}^{\dagger}\hat{a}_{\ell'}\rangle$. 

\begin{figure}
    \centering\includegraphics[width=0.95\linewidth]{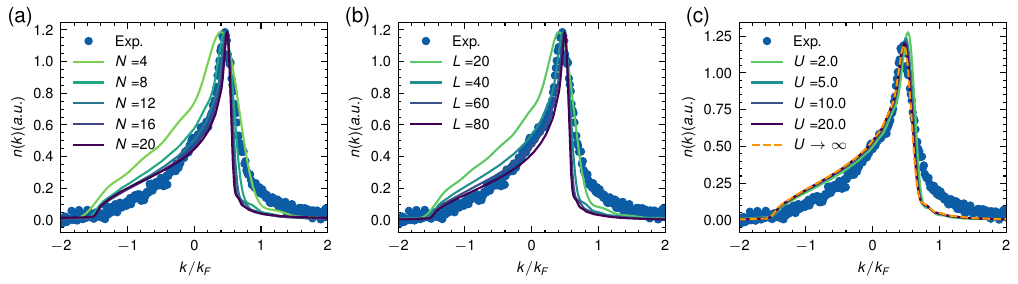} %
    \caption{\textbf{Benchmarking the experimental data by the AHM.} (a) Effect of the filling factor at $L=40$ and $U\rightarrow\infty$. (b) Effect of the system size at filling $1/4$ at $U\rightarrow\infty$. (c) Effect of the on-site interaction $U$ at $L=40, N=10$. The blue dots are the experimental data for $\theta/\pi=0.53(2)$; the solid lines are the prediction from the AHM. Note that in (a,b), the peak has been rescaled to match the amplitude of the experimental data.}
    \label{fig_AHM_para}
\end{figure}

To benchmark the anyonic correlations from the AHM with the experiment, we consider the experimental impurity momentum distribution at $\theta/\pi=0.53(2)$ as an example. As shown in Fig.~\ref{fig_AHM_para}, the agreement between the experimental data and the anyonic momentum distribution obtained from the AHM improves for large system sizes and low filling.

\section{Ground-state properties of the swap model $\hat{H}_{\text{Swap}}$}

\begin{figure}
    \centering\includegraphics[width=0.95\linewidth]{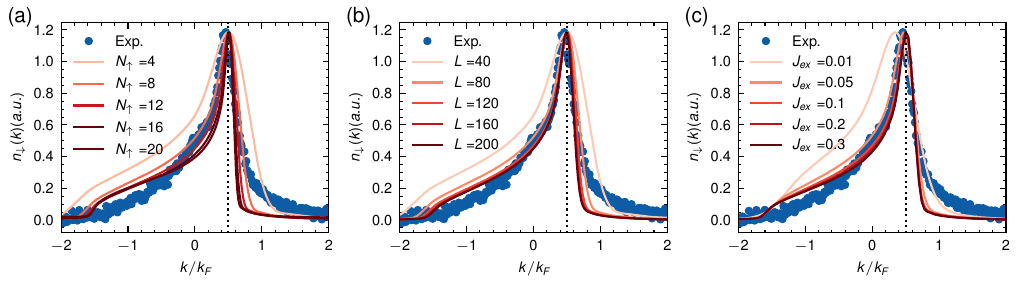} %
    \caption{ \textbf{Effect of different system parameters of the swap model.} (a) Quasi-momentum distribution for the impurity for varying number of host particles $N_{\uparrow}$ with $J_\text{ex}=0.1, L=40$; (b) Quasi-momentum distribution of impurity for varying system size $L$ with $J_\text{ex}=0.1, N_{\uparrow}/L=0.1$; (c) Quasi-momentum distribution of impurity for varying $J_{\text{ex}}$ with $N_{\uparrow}=12, L=120$. The black dotted lines in (a-c) show the location of the peak, which is given by $k_{\text{peak}}/k_{\text{F}}=\theta/\pi$. Here the quasi momenta are plotted in the unit of $k_{\text{F}}=\pi N_{\uparrow}/L$. Note that the peak has been rescaled according to the experimental data. } 
    \label{fig_swap_para}
\end{figure}

Next, we investigate the effect of various system parameters of the swap model, specifically the number of host particles $N_{\uparrow}$, the length of the system chain $L$, and the swapping strength $J_{\text{ex}}$. As shown in Fig.~\ref{fig_swap_para}, changes in these parameters lead to similar behavior, in the sense that larger values give rise to a narrower quasi-momentum distribution (less uncertainty in momentum space). Note that the height of the theoretical quasi-momentum distribution results are rescaled by the experimental data. 

\begin{figure}
    \centering\includegraphics[width=0.8\linewidth]{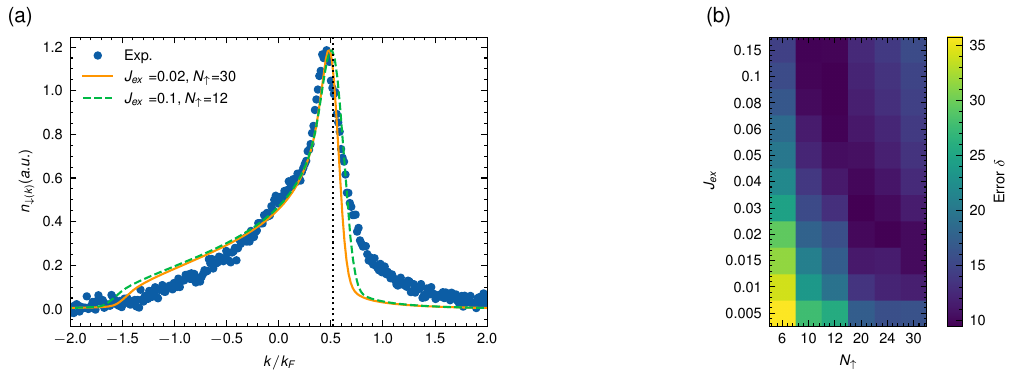} %
    \caption{ \textbf{Comparison between experiment and the predictions of the swap model for different values of $J_\text{ex}$ and $N_\uparrow$.} (a) Comparison between experimental data (blue dots) and the quasi-momentum distribution obtained by using the swap model. (b) Interpolation error as a function of $J_{\text{ex}}$ and $N_{\uparrow}$. } 
    \label{fig_swap_exp}
\end{figure}

Now we fix the system length to $L=120$, and tune the parameters $J_{\text{ex}}$ and $N_{\uparrow}$. By computing the residuals $\delta$ from the experimental data, one can see that low values of $\delta$ appear in a wide parameter regime, as shown in Fig.~\ref{fig_swap_exp}(b). Similar results can be obtained by using either low filling (e.g. $n=0.1$) with $J_{\text{ex}}=0.1$ or higher filling (e.g.\ $n=0.25$) but lower $J_{\text{ex}}=0.02$, see Fig.~\ref{fig_swap_exp}(a). This trend indicates that in the thermodynamic limit ($N_\uparrow\rightarrow\infty$), one would have $J_{\text{ex}}\rightarrow 0$. We can conclude that the behavior of anyionization in a finite system with open boundary conditions is well captured by using a finite $J_{\text{ex}}$ in our swap model.

\end{document}